\documentclass[twocolumn,showkeys,preprintnumbers,amsmath,amssymb]{revtex4}

\usepackage{graphicx}
\usepackage{dcolumn}
\usepackage{bm}
\usepackage{subeqnarray}

\def \kmsmpc{${\rm ~km ~s^{-1}Mpc^{-1}}$}

\date{\today }

\begin{document}

\title{On the anomalous acceleration of Pioneer spacecraft}

\author{Moshe Carmeli}
 \email{carmelim@bgu.ac.il}
\affiliation{Department of Physics, Ben Gurion University of the Negev, \\ Beer Sheva
84105, Israel}

\author{John G. Hartnett}
 \email{john@physics.uwa.edu.au}
\affiliation{School of Physics, the University of Western Australia,\\35 Stirling Hwy, Crawley 6009 WA Australia}

\author{Firmin J. Oliveira}
 \email{f.oliveira@jach.hawaii.edu}
\affiliation{Joint Astronomy Centre, 660 N. A'ohoku Place, \\University Park, Hilo, Hawai'i, U.S.A.}

\begin{abstract}
The anomalous acceleration of Pioneer 10 and 11 spacecraft of $(8.74 \pm 1.33) \times 10^{-8}\;cm. s^{-2}$ fits with a theoretical prediction of a minimal acceleration in nature of about $7.61 \times 10^{-8}\;cm. s^{-2}$
\end{abstract} 


\maketitle

\section{\label{sec:Introduction} \bf{Introduction}}

It is well known that both the Pioneer 10 and 11 spacecraft experience an anomalous acceleration toward the Sun, amounting to $(8.74\pm 1.33)\times 10^{-8}cm/s^2$. The anomaly has been found in the Pioneer spacecraft data as they traversed distances from 20 AU to 70 AU since 1987. This has been reported by Anderson {\it et al.} \cite{Anderson2002}, but it has remained unaccounted for. 

When a particle is located in an empty space its motion is determined by the
forces acting on it and if there are no forces the particle will stay in its location
or move with a constant velocity. If the particle is located in a moving medium
like a river its velocity with respect to an observer standing on the bank of the
river is the sum of the velocity of the particle with respect to water and the
speed of the water with respect to the observer. This is the situation if the
water flows with a constant velocity both in time and location. The situation in
the cosmos, even if there are no gravitational forces, is a bit more complicated.
Because of the Hubble expansion a particle will move along with it. But now
the expansion depends on the location as is well known and thus the motion
of the particle will experience an additional acceleration. This effect can be
understood intuitively by the relation $t = x/v = dx/dv = v/a$, where $a$ is the
acceleration. If we go to the maximum values of this equation, and write for
$t_{max} = \tau$ as the maximum time allowed in nature and noting that $v_{max} = c$ we
then have $t_{max} = c/a_{min}$. Hence $a_{min} = c/\tau$. This calculation can be obtained
in a rigorous way from the cosmological transformation that relates space, time
and redshift expansion, which is an extension of the Lorentz transformation keeping with cosmological invariance (Carmeli \cite{Carmeli1995,Carmeli1996}). The value of the constant
$\tau$ can be determined from the cosmological constants when gravity is invoked
and found to be $\tau \approx 12.486\;Gyr$ (Ref. \cite{Carmeli2002}, Eq. (A.66), p. 138).

The minimal acceleration in nature is directed, locally, along the motion
of the particle. This follows from the cosmological principle whereby any point
in the Universe can be considered as the center from which the Hubble flows
locally in all outer directions. It is as if the particle were on top of a hill and was
released to fall freely, although the analogy is far from being accurate since the
minimal acceleration in nature is a kinematical quantity, whereas the latter case
is due to gravity. The minimal acceleration acts as a rocket, causing an increase
in the particle's velocity $v = a_{min}t$ and in terms of distance $d = a_{min}t^2/2$.
In this paper we give the essentials that lead to a minimal acceleration in
nature. In Section \ref{sec:costrans} we give the cosmological transformation. In Section \ref{sec:tau} we relate that transformation to measurements from recent experiments. In the last
section we give some concluding remarks.

\section{\label{sec:costrans} \bf{The Cosmological Transformation}}

The Universe expands according to Hubble's law $R = H_0^{-1}v$, where $H_0$ is the
Hubble parameter (usually called the Hubble constant). $H_0$ is not a constant;
it gives the expansion rate at a certain cosmic time. $H_0^{-1}$ is usually called the
Hubble time and it gives a measure of the age of the Universe. However it is not really
the age of the Universe. All that is valid assuming the Universe has a structure
with gravitation. Now we ask, what will happen if there was an abstract
Universe with no gravity? Obviously, such a Universe does not exist in nature.
Nevertheless, it is useful to consider such an abstract Universe for pedagogical
reasons. In such an abstract Universe one has to replace the Hubble parameter
$H_0$ by a constant and that Universe will expand according to $R = \tau v$, where $\tau$
is the value of $H_0^{-1}$ in the limit of zero gravity.

Since the Universe extends in three spatial dimensions, the above relation of
the expansion in zero gravity will have the form $(x^2 + y^2 + z^2)^{1/2} = \tau v$, or
\begin{equation} \label{eqn:Eqn1}
x^2 + y^2 + z^2-\tau^2 v^2 = 0 
\end{equation}
We thus obtain a new kind of line element given by
\begin{equation} \label{eqn:Eqn2}
ds^2=\tau^2 v^2-(x^2 + y^2 + z^2);
\end{equation}
$ds^2$ is obviously zero when there is an expansion in a radial direction, but it
is otherwise not zero. Hence such an abstract Universe expands according to
the null condition of the line element. It remains to determine the constant $\tau$
and to relate it to measured quantities in cosmology. 

For a Universe filled with gravity it will be assumed that there is a generalization to the above line
element without gravity in the form of
\begin{equation} \label{eqn:Eqn3}
ds^2=g_{\mu \nu}dx^{\mu}dx^{\nu},
\end{equation}
where $g_{\mu \nu}$ is a four-dimensional metric that should be determined by the Einstein
field equations, and the coordinates are $x^0 = \tau v$, $x^k$, $k=1,2,3$, are the
spatial coordinates. This is not the standard Einstein theory of gravitation in
spacetime. Rather, it is a distribution theory in space and velocity at a very
definite time. The points $x, y, z$  $(x^k)$ give the location of galaxies in space,
$(x^2 + y^2 + z^2)^{1/2}$ is the radial distance of the galaxy from the observer, and $v$
is the outward radial velocity of the galaxy. Accordingly, each galaxy in the
Universe is presented in the four-dimensional curved manifold which has to be
determined by solving the field equations.

The cosmological transformation is the analogue of the familiar Lorentz
transformation in spacetime, but now the new transformation relates space coordinates
to the outward velocity of the expansion. For simplicity we assume
that the expansion is along the $x$ axis. Hence Hubble's law in the stationary
system $K$ and the transformed system $K'$ is given by
\begin{equation} \label{eqn:Eqn4}
x = \tau v \, , \, x' = \tau v',
\end{equation}
where $x, v$ and $x', v'$ are measured in $K$ and $K'$ respectively. If we assume that $x, v$ and
$x', v'$ transform linearly,
\begin{subeqnarray} 
x' = ax - bv \label{eqn:Eqn5a},\\
x = ax' + bv' \label{eqn:Eqn5b},
\end{subeqnarray}
where $a$ and $b$ are some variables that are independent of the coordinates. At
$x' = 0$ and $x = 0$, Eqs. (5) yield, respectively,
\begin{subeqnarray} 
b/a = x/v = t, \label{eqn:Eqn6a} \\
b/a = -x'/v' = t. \label{eqn:Eqn6b}
\end{subeqnarray}

From Eq. (\ref{eqn:Eqn6a}) we have
\begin{equation} \label{eqn:Eqn7}
t = x/v = dx/dv = v/a,
\end{equation}
where $a$ is the acceleration. At the maximum value of the above equation one obtains
\begin{equation} \label{eqn:Eqn8}
t_{max} = \tau = (v/a)_{max} = c/a_{min}.
\end{equation}
It thus appears that in nature there is a minimal acceleration given by
\begin{equation} \label{eqn:Eqn9}
a_{min} = c/\tau \approx 10^{-8} cm/s^2.
\end{equation}

The form of the line element $ds^2$ given in Eq. (\ref{eqn:Eqn2}) suggests that the new transformation
will have the form
\begin{equation} \label{eqn:Eqn10}
x' = \frac{x- tv}{(1-t^2/\tau^2)^{1/2}} \, , \, v' = \frac{v- xt/\tau^2}{(1-t^2/\tau^2)^{1/2}},
\end{equation}
$y' = y$, $z' = z$. In the above equations $t$ is the cosmic time measured backward
($t = 0$ now), and we see that $t/\tau$ replaces the familiar factor $v/c$ of special
relativity. Thus $\tau$ takes the role of the speed of light $c$ of special relativity
theory. Of course one can rewrite the cosmological transformation in terms of
the ordinary cosmic time that is equal to zero at the Big Bang. Notice that as the
cosmic time $t$ approaches $\tau$ the denominators in the cosmological transformation
(\ref{eqn:Eqn10}) become very small just like the situation in the Lorentz transformation. It
thus appears that $\tau$ plays the role of a maximum time in nature or, in other
words, it is the Big Bang time just like $c$ is the maximum velocity in special
relativity theory.

In the next section we relate the constant $\tau$ to some constants known in
cosmology and thus determine its value.

\section{\label{sec:tau} Determining the value of $\tau$}

After solving the Einstein field equations one finds (\cite{Carmeli2002}, Appendix A)
\begin{equation} \label{eqn:Eqn11}
r = \tau v [1 + (1 -\Omega) v^2/6c^2].
\end{equation}
The above equation gives the relationship between the distance of a galaxy and
its velocity. Inverting the above equation by writing it as $v = H_0r$, we obtain
in the lowest approximation the value of the Hubble parameter $H_0$,
\begin{equation} \label{eqn:Eqn12}
H_0 = h [1 + (1 -\Omega) v^2/6c^2],
\end{equation}
where $h = 1/\tau$, and $\Omega$ is the matter density in the Universe. The above equation
can also be written in the form
\begin{equation} \label{eqn:Eqn13}
H_0 = h [1 + (1 -\Omega) z^2/6],
\end{equation}
where $z$ is the redshift and $\Omega = \rho_m/\rho_c$ with $\rho_c = 3h^2/8\pi G$ and $G$ is Newton's gravitational constant. It will be noted that $\rho_c \approx 1.194 \times 10^{-29}\; g/cm^2$; it is different from the
standard critical density $\rho_c$ defined by $\rho_c = 3H^2_0/8\pi G$. If we take for simplicity
$z = 1$ and $\Omega = 0.245$ the last equation then gives
\begin{equation} \label{eqn:Eqn14}
H_0 = 0.874h.
\end{equation}
We assume that $H_0 = 70$ \kmsmpc, hence $h = (70/0.874)$  \kmsmpc  or
\begin{equation} \label{eqn:Eqn15}
h = 80.092 \; \rm ~km ~s^{-1}Mpc^{-1}
\end{equation}
and thus $\tau$ will have the value
\begin{equation} \label{eqn:Eqn16}
\tau = 12.486 \; Gyr = 3.938 \times 10^{17}\; s.
\end{equation}

From the above one finds that the minimal acceleration in nature is given
by $a_{min} = c/\tau \approx 3\times 10^{10}/3.938 \times 10^{17} = 7.61 \times 10^{-8}\; cm/s^2$.

\section{\label{sec:Conclude} Concluding remarks}

The application of the cosmological transformation, and hence cosmological
relativity, seems to explain the Pioneer anomaly in a striking agreement. One
wonders, however, whether this modification of Einstein's theory due to the
Hubble expansion shouldn't have the same effect on the planets of our solar
system as well? The answer is no. That is like asking whether the Lorentz
transformation is applicable to the electrons in the atom. In the latter case
the mass of the electron as well as its speed around the nucleus of the atom is
changing extremely rapidly. Classical arguments seem not to be applicable here.
In the cosmological case a similar situation holds, for different physical reasons,
since the planets and the Sun are `tied' by their gravitational attraction, and
provide a closed system.


\begin{thebibliography}{99}
\bibitem{Anderson2002} Anderson, J.D. \textit{et al.}, \textit{Physical Review D} \textbf{65}, 082004 (2002)
\bibitem{Carmeli1995} M. Carmeli, ``Cosmological special relativity: a special relativity for cosmology,'' \textit{Found. Phys} \textbf{25} 1029 (1995)
\bibitem{Carmeli1996} M. Carmeli, ``Cosmological special relativity,'' \textit{Found. Phys} \textbf{26} 413 (1996)
\bibitem{Carmeli2002} Carmeli, M., \textit{Cosmological Special Relativity,} Second Edition (World Scientific, Singapore, 2002)

\end{thebibliography}
\end{document}